\documentclass[twocolumn,showpacs,preprintnumbers,amsmath,amssymb,prl]{revtex4-1}

\usepackage{dcolumn}
\usepackage{bm}

\usepackage{color}

\usepackage{graphicx}
\usepackage{natbib}
\usepackage{subfigure}
\usepackage{cases}
\usepackage{epstopdf}

\newcommand{\beq}{\begin{equation}}
\newcommand{\eeq}{\end{equation}}
\newcommand{\curl}{\nabla\times}

\begin{document}
\def\bfB{\mbox{\bf B}}
\def\bfQ{\mbox{\bf Q}}
\def\bfD{\mbox{\bf D}}
\def\etal{\mbox{\it et al}}
%
\title{On the edge of an inverse cascade}
\author{Kannabiran Seshasayanan$^1$, Santiago Jose Benavides$^2$, Alexandros Alexakis$^1$}

\affiliation{
$^1$Laboratoire de Physique Statistique, CNRS UMR 8550,  \'Ecole Normale Sup\'erieure,  Paris, France, 
CNRS, Universit\'e Pierre et Mari\'e Curie, Universit\'e Paris Diderot, 24 rue Lhomond, 75005 Paris, France\\ 
$^2$Physics Department, The University of Texas at Austin, Austin, TX 78712, USA}

\date{\today}

\begin{abstract}
We demonstrate that systems with a parameter-controlled inverse cascade can exhibit critical behavior
for which at the critical value of the control parameter the inverse cascade stops.
In the vicinity of such a critical point standard phenomenological estimates for the energy balance 
will fail since the energy flux towards large length scales becomes zero.
We demonstrate these concepts using the computationally tractable model of two-dimensional magneto-hydrodynamics in a periodic box.
In the absence of any external magnetic forcing the system reduces to 
hydrodynamic fluid turbulence with an inverse energy cascade. In the presence of strong magnetic forcing 
the system behaves as 2D magneto-hydrodynamic turbulence with forward energy cascade. 
As the amplitude of the magnetic forcing is varied
a critical value is met for which the energy flux towards the large scales becomes zero.
Close to this point the energy flux scales as a power law with the departure from the critical point and the normalized amplitude of the fluctuations diverges. 
Similar behavior is observed for the flux of the square vector potential for which no inverse flux is observed for weak magnetic forcing, while 
a finite inverse flux is observed for magnetic forcing above the critical point. 
We conjecture that this behavior is generic for systems  of variable inverse cascade.

\end{abstract} 

\maketitle
In many dynamical systems in nature energy is transfered to smaller or to larger length scales
by a mechanism known as forward or inverse cascade, respectively.
In three-dimensional hydrodynamic (HD) turbulence energy cascades forward from large to small scales while in two-dimensional  HD
turbulence energy cascades inversely from small scales to large scales \cite{Turb,2DT}. When the  dissipation coefficients 
are very small (large Reynolds numbers) the rate that energy is dissipated $\epsilon$ equals the flux of energy $\Pi_{_E}$ introduced
by the cascade and thus this process is of fundamental interest for many fields (astrophysics, atmospheric sciences, industry, ect.). 
There are some examples, however, that have a mixed behavior such as fast rotating fluids, 
stratified flows, conducting fluids in the presence of strong
magnetic fields, or flows in constrained geometry \cite{Rot,strat,geophys,B0,B1,2DH}. In these examples the injected energy cascades both 
forward and inversely in fractions that depend on the value of a control parameter $\mu$ (rotation rate/magnetic field/aspect ratio).
In rotating flows, for example, when the rotation is weak the behavior of the flow is similar to isotropic turbulence and
energy cascades forward. As the rotation rate is increased variations along the direction of rotation are suppressed
and the flow starts to become quasi-2D. Eventually when rotation is strong enough the two-dimensional component of the flow 
dominates and energy starts to cascade inversely to the large scales. 
This dual cascade behavior is not restricted to quasi-2D flows neither to the cascade of energy.
It is also observed in wave systems such as surface waves \cite{Lvov}, 
elastic waves \cite{elast} 
and quantum fluids \cite{qfluid}.
The variance of a passive scalar is also shown to display a change of direction of cascade when the compressibility of the flow,
or its geometry vary \cite{scal,2DH}. Finally, in turbulent dynamo flows the flux of magnetic helicity changes direction 
depending on the sign of the kinetic helicity \cite{Mhel1,Mhel2} with no inverse cascade of magnetic helicity observed for non-helical flows. 
Thus, the transition from forward to inverse cascade by the variation of a control parameter $\mu$ seems to be a common property for many 
out of equilibrium systems.

This transition can occur either in a smooth way or by a bifurcation at a critical value $\mu_c$ of the control parameter for which the transition 
from forward to inverse cascade begins. Such a transition differs from regular  scenario of bifurcation
from laminar to turbulent flows since in this case the system transitions from one fully turbulent 
state to another fully turbulent state, making regular expansions non-applicable.
This behavior resembles more phase transitions in equilibrium statistical mechanics where 
an order parameter  (for instance the magnetization for a system of spins) deviates continuously from zero but with
discontinuous derivatives and its susceptibility theoretically diverging at the critical point.
The order parameter then depends on the distance from the critical point as a power-law.  
The exponents of these power-laws are referred to as the critical exponents of the system.
Following this analogy we also expect for the out of equilibrium systems 
that the energy flux of the inverse cascade will depend on the
distance from the critical point $\mu_c$ as a power-law $\Pi_E \propto (\mu-\mu_c)^\gamma$.

The existence of such a critical point then forces us to rethink some of the fundamental 
relations of turbulence in its vicinity.
In the limit of small dissipation coefficients (viscosity, hypo-viscosity etc.)
the relation for the energy flux $\Pi_{_E} = C_{_K} U^3/L$ (where $U$ is the rms velocity and $L$ is the 
integral length scale and $C_{_K}$ is a constant that can be measured from experiments.) 
can be derived by simple dimensional arguments.  The same relation holds for both inverse (2D) and forward (3D) cascades
but with different values of $C_{_K}$. In the scenario of a cascade transition the parameter $C_{_K}$ 
is a function of $\mu$ that varies from zero to an order one value. This relation is non-trivial and cannot be predicted 
by dimensional arguments and thus not even simple estimates of order of magnitude can be made for the dissipation rate in these systems.

In this work we try to demonstrate these ideas for two-dimensional incompressible
magneto-hydrodynamics (MHD) in a double periodic square domain of size $2\pi L$. 
As we describe below, this system has a variable inverse cascade of energy. Studying this system also has the advantage
that it can be studied at low computational cost due to its low dimensionality.
The dynamical equations for the system can be written in terms of the vorticity $w = {\bf e}_z\cdot \curl {\bf u}$ (where ${\bf u}$ is the velocity field) and
the vector potential $a$  of the magnetic field  ${\bf b }=\curl ({\bf e}_z a)$. They are given by:
\begin{eqnarray}
\partial_t w  + {\bf u \cdot \nabla} w  =&  {\bf b \cdot \nabla} j  &+ \nu^+ \Delta^{n} w  + \nu^-  \Delta^{-m} w  + \phi_w\nonumber \\
\partial_t a  + {\bf u \cdot \nabla} a  =&                          &+ \eta^+\Delta^{n} a  + \eta^- \Delta^{-m} a  + \phi_a.
\end{eqnarray}
where ${\bf e}_z$ is the unit vector normal to the plane, and $j={\bf e}_z\cdot \curl {\bf b}$.
$\phi_w$ and $\phi_a$ introduce the mechanical force ${\bf F_u}=-\Delta^{-1} \curl ({\bf e}_z \phi_w)$ 
                            and the magnetic   force ${\bf F_b}= \curl ({\bf e}_z \phi_a)$
that inject energy in the system at the scale $k_f^{-1}$. 
In particular we have chosen $\phi_w=2f_0k_f \cos(k_fx)\cos(k_fy)$ and $\phi_a=\mu f_0 k_f^{-1} \sin(k_fx)\sin(k_fy)$.
Therefore $\mu=\|{\bf F_b}\|/\|{\bf F_u}\|$  expresses the ratio of magnetic to mechanical forcing.
Energy is removed from the system by the terms proportional to $\nu^+$ and $\eta^+$ in the small scales 
and by $\nu^-$ and $\eta^-$ in the large scales. The indexes $m,n$ give the order of the laplacian used. The physically motivated values are
$n=1$ and $m=0$, however, to obtain a larger inertial range we chose $n=m=2$. For all runs we have fixed $\nu^+=\eta^+$ and  $\nu^-=\eta^-$.

With these choices we are left with four control parameters. We have 
a Reynolds number for the forward energy cascade: $Re^+\equiv (f_0^{1/2}k_f^{1/2-2n})/\nu^+$, 
a Reynolds number for the inverse energy cascade: $Re^-\equiv (f_0^{1/2}k_f^{1/2+2m})/\nu^-$, 
the ratio of the forcing length scale to the box size $k_fL$ and $\mu=\|{\bf F_b}\|/\|{\bf F_u}\|$.
The last parameter controls the transition from an inverse cascade to a direct cascade.

The system in the absence of forcing and dissipation conserves two positive-definite quadratic quantities: the total energy 
$E=\frac{1}{2}\langle {\bf u^2+b^2} \rangle  $
and the square vector potential
$A =\frac{1}{2} \langle a^2 \rangle$ where $\langle\cdot \rangle$ indicates spacial average.  
In the absence of any external magnetic field or a magnetic source $\phi_a$ any
magnetic field fluctuations that exist at t=0 will die out (due to the
anti-dynamo theorem of 2D flows \cite{ADT}) and the system will reduce to ordinary 2D
fluid turbulence with an inverse cascade for energy \cite{2DT}
and a forward cascade of $A$ that acts like the variance of a passive scalar \cite{batch}. 
If, however, a magnetic force ${\bf F_b}$  exists (and is sufficiently strong) the flow will sustain magnetic field 
fluctuations and become magnetic dominated with a forward energy cascade \cite{MHDT} and an inverse cascade of A \cite{2DMHD1,2DMHD2}.

The flux of energy at any wavenumber $k$ is defined as
\beq                                                                                                                                                      
\Pi_{_E}(k) \equiv \langle {\bf  u}_k^< {\bf   ( u\cdot\nabla u  - b\cdot\nabla b)  +  b}_k^< {\bf ( u\cdot\nabla b - b\cdot\nabla u)}\rangle 
\eeq
while
$
\Pi_{_A}(k) \equiv \langle a_k^<  ( {\bf u}\cdot\nabla a) \rangle 
$
defines the flux of the square vector potential \cite{MHDT}. Here $f_k^<$ represents
the filtered field $f$ so that only the Fourier modes ${\bf k}$ satisfying $|{\bf k}|\le k$ have been kept.
The dissipation rate of energy at the small scales is defined in terms of the Fourier components of
the two fields ${\bf \tilde{u}_k,\tilde{b}_k }$ as:
\beq                                                                                                                                                      
\epsilon_{_E}^+ \equiv \nu^+\sum_{\bf k\ne0} |{\bf k}|^{2n} (|{\bf \tilde{u}_k}|^2+|{\bf \tilde{b}_k}|^2) ,
\eeq
while in the large scales as
\beq                                                                                                                                                      
\epsilon_{_E}^- \equiv \nu^- \sum_{\bf k\ne0} |{\bf k}|^{-2m} (|{\bf \tilde{u}_k}|^2+|{\bf \tilde{b}_k}|^2) .
\eeq
Similarly we define the dissipation of $A$:
\beq                                                                                                                                                      
\epsilon_{_A}^+ \equiv \nu^+\sum_{\bf k\ne0} |{\bf k}|^{ 2n} |\tilde{a}_{\bf k}|^2 \quad \mathrm{and} \quad  
\epsilon_{_A}^- \equiv \nu^-\sum_{\bf k\ne0} |{\bf k}|^{-2m} |\tilde{a}_{\bf k}|^2.
\eeq
The dissipation rates at the large scales $\epsilon_{_E}^-$ and $\epsilon_{_A}^-$ 
provide a measure of the strength of the inverse cascade. In the infinite $Re^-$ limit
$\epsilon_{_E}^-$ and $\epsilon_{_A}^-$ will be non-zero if and only if an inverse cascade exists.
For any finite value of $Re^-$, however, some weak large scale dissipation will exist due to
the finite value of the dissipation coefficients $\nu^-,\eta^-$. To investigate the 
large $Re^-$ limit the following procedure was followed: for fixed $Re^\pm$ and $k_fL$
simulations  were performed for different values of $\mu$ varying from 0 to 1. 
The flow was simulated using a standard pseudospectral code \cite{Gomez}.
The simulations were run long enough in time so that a steady state is reached and long time averages can be performed.
Then keeping $Re^+$ fixed we increased $k_fL$ and $Re^-$ and a new series of simulations was performed
varying $\mu$ in the same range. That way the inertial range of the inverse cascade was increased
and the large $Re^-$ limit was  approached. This way we could check to what extend the obtained results converged to
an $Re^-$ independent behavior. The choices for $k_fL$ were $k_fL=8,16,32,64$, and resolutions
of up to $4096^2$ grid points were used. $Re^-$ was chosen as large as possible so that a clean 
inertial range is obtained, but sufficiently small so that no large scale condensate formed at the large scales.

\begin{figure}                                                                                                                %
\includegraphics[width=8.0cm]{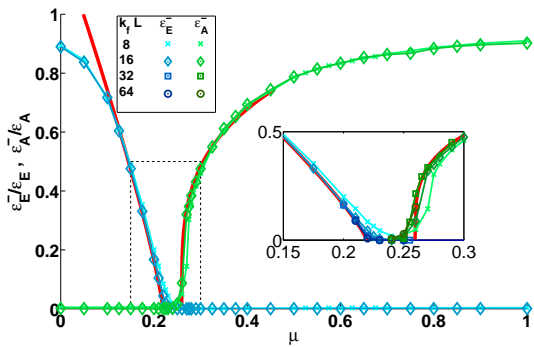}                                                                                    %
\caption{\label{fig1} Normalized energy dissipation in the large scales $\epsilon_{_E}^-/\epsilon_{_E}$                       %
                 and  normalized square vector potential dissipation    $\epsilon_{_A}^-/\epsilon_{_A}$                       %
                 as a function of $\mu$ for different values of $k_fL$ (as indicated by the table in the figure).             %
                 The inset shows a zoom of the same data close to the critical point. The red lines show the fitting          %
                 curves $(\mu_c-\mu)^{\gamma_{_E}}$ and $(\mu-\mu_c)^{\gamma_{_A}}$. }                                        %
\end{figure}                                                                                                                  %
In figure \ref{fig1} we present the large scale dissipation rates $\epsilon_{_E}^-$ and $\epsilon_{_A}^-$
normalized by the total injection rates 
$\epsilon_{_E}=\epsilon_{_E}^-+\epsilon_{_E}^+$ and 
$\epsilon_{_A}=\epsilon_{_A}^-+\epsilon_{_A}^+$ as a function of $\mu$.  
The different {\it colors/symbols} indicate the the different values of $k_fL$ used.
For small values of $\mu$ the system behaves like an HD flow with an inverse cascade of energy 
indicated by the fact that almost all of the injected energy is dissipated in the large scales $\epsilon_{_E}^-/\epsilon_{_E}\simeq 1$.
At the same time no inverse cascade of $A$ is observed since $\epsilon_{_A}^-/\epsilon_{_A}\simeq 0$.
For $\mu\simeq 1$, on the other hand, the system behaves like an MHD flow with no inverse cascade of energy 
$\epsilon_{_E}^-/\epsilon_{_E}\simeq 0$ but an inverse cascade of $A$, ($\epsilon_{_A}^-/\epsilon_{_A}\simeq 1)$. 
For intermediate values of $\mu$ energy and square vector potential are dissipated both at large and small scales 
at fractions that depend on $\mu$ with the inverse cascade of energy decreasing with $\mu$ and the inverse cascade of
$A$ increasing with $\mu$.

Around $\mu \simeq 0.22$ the inverse cascade of energy ends and at $\mu\simeq 0.25$ the inverse cascade of $A$ begins. 
The two cascades thus seem to have different critical values.
The transition is less sharp for small values of $k_fL$.
However as the domain size $k_fL$ is increased the curves converge to a sharp transition and 
the two transition points approach each other.
This can be seen more clearly at the inset where a close up at the critical point is shown.
Due to the long time to reach saturation in the presence of inverse cascades, the large 
$k_f L$ cases (that also required the largest resolutions) were limited only to  values of $\mu$ 
close to the critical points.

Both dissipation rates scale as a power laws 
$\epsilon_{_E}^-\propto (\mu_c-\mu)^{\gamma_{E}} $ and $\epsilon_{_A}^-\propto (\mu-\mu_c)^{\gamma_{A}}$,
with a best fit leading to $\gamma_{E}\simeq 0.82$ and $\gamma_{A}\simeq 0.27$.  
However, not being able to determine precisely the value
of $\mu_c$ significantly limits the accuracy of these measurements. Furthermore the exact location of $\mu_c$  
was also found to depend on the value of $Re^+$ caused by the dependence of magnetic energy on $Re^+$.
Besides the energy flux other quantities also showed power law behaviors. A full presentation of the results of 
the 2D-MHD  system however will be reported in a lengthier report. Here we focus on the more general features of the 
cascade transition scenario. 

\begin{figure}                                                                                                                %
\includegraphics[width=8.0cm]{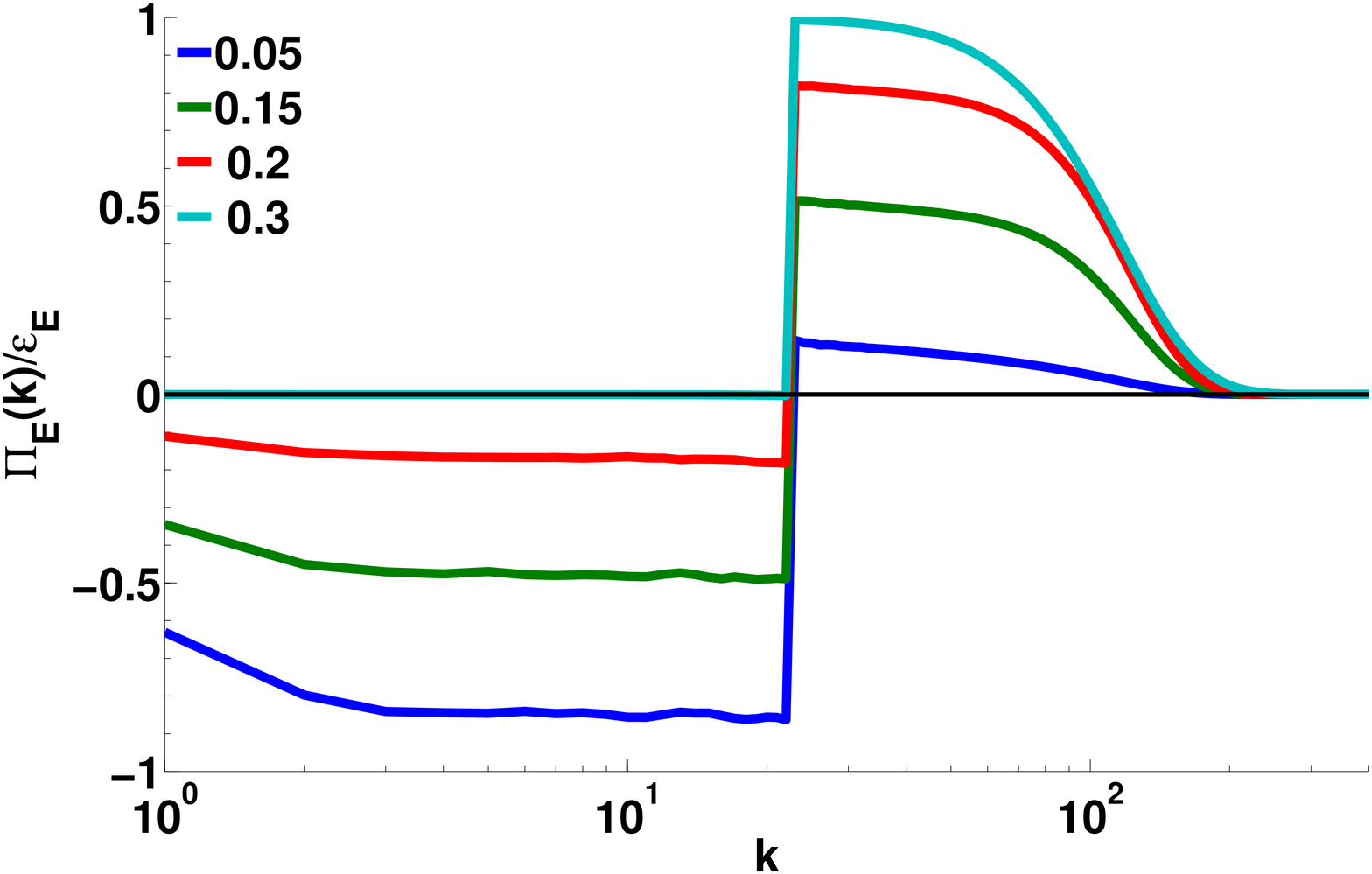}                                                                                    %
\includegraphics[width=8.0cm]{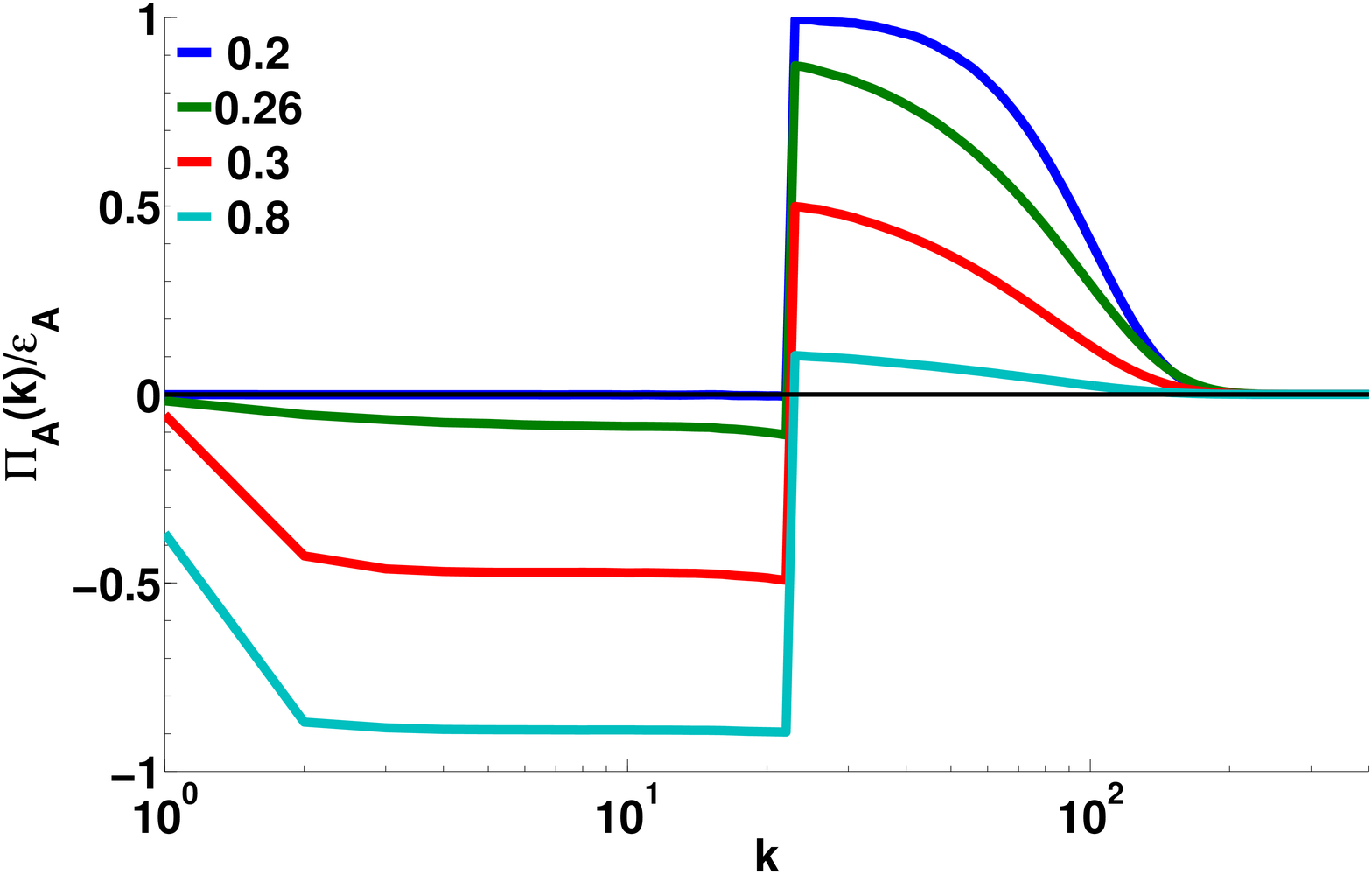}                                                                                    %
\caption{\label{fig2} The energy flux $\Pi_{_E}(k)$ (top panel) and the flux of the square vector potential $\Pi_{_A}(k)$     %
                      (lower panel) for different values of $\mu$ around the critical point and $k_fL=16$. }                  %
\end{figure}                                                                                                                  %
The cascade transition 
can be seen in the plots of the two normalized fluxes $\Pi_{_E}(k)/\epsilon_{_E}$ and  $\Pi_{_A}(k)/\epsilon_{_A}$ 
shown for various values of $\mu$ in figure $\ref{fig2}$. The two fluxes remain constant in $k$ for a wide range of wavenumbers. 
As the parameter $\mu$ is varied
the flux varies from $0$ to $-1$ (in the range $k<k_f)$ and from $0$ to $+1$ (in the range $k>k_f)$. 
At intermediate values of $0<\mu<\mu_c$ part of the energy cascades 
to small scales and part to the large scales and similarly for $A$ in the range
$\mu>\mu_c$.
 
\begin{figure}                                                                                                                %
\includegraphics[width=8.0cm]{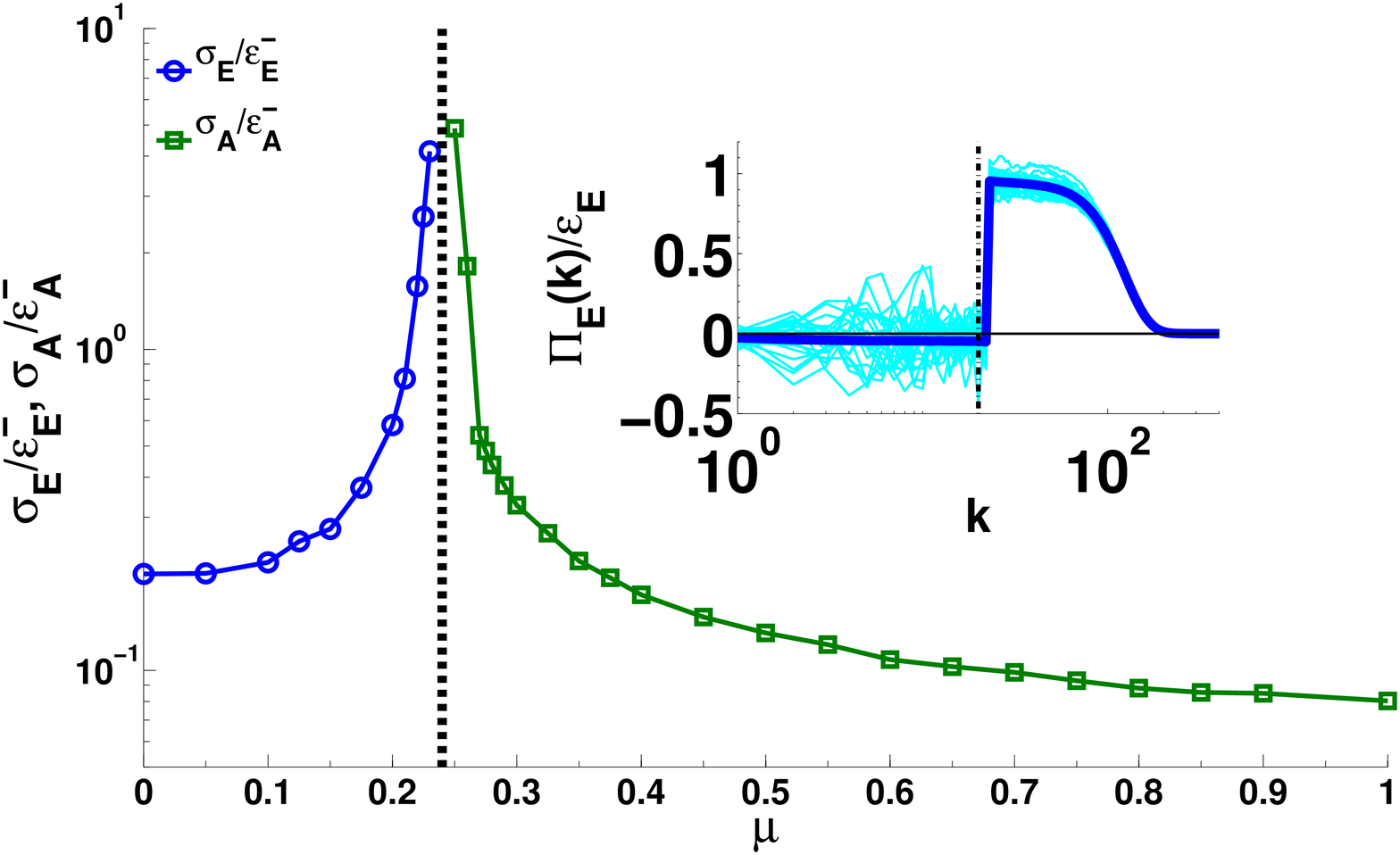}                                                                                    %
\caption{\label{fig3} The inset     shows the instantaneous energy flux $\Pi_{_E}(t,k)$  with cyan lines for various times    %
                    $t$ and the time averaged energy flux $\Pi_{_E}(k)$ with the dark blue line for $\mu=0.21$ and $k_fL=16$. %
                      The main figure shows the                                                                               %
                      variance of the energy flux $\sigma_{_E}=\langle (\Pi_{_E}(t,k)-\Pi_{_E}(k))^2\rangle^{1/2}$            %
                      and the variance of the square vector potential flux $\sigma_{_A}$ normalized by the large scale        %
                      dissipations.                                                                                           %
                      The variance was evaluated at $k$ slightly smaller than the forcing wavenumber indicated by the         %
                      vertical dashed line in the inset.    }                                                                 %
\end{figure}

We emphasize the role of the flux fluctuations close to the critical value.
The inset in figure \ref{fig3} shows the time averaged flux $\Pi_{_E}(k)$ with the dark line (blue online) while with the 
light gray (cyan online) lines the instantaneous flux $\Pi_{_E}(t,k)$ is plotted at various instances of time. 
The instantaneous flux is not constant in $k$, on the contrary it fluctuates taking both positive and negative values. 
The amplitude of the fluctuations of the flux exceed by much the averaged flux. In fact as figure 
\ref{fig3} shows the relative amplitude of the fluctuations $\sigma_{_E}$ and $\sigma_{_A}$
(standard deviation from the mean value) to the averaged value diverges as the critical point
is approached. 

\begin{figure}                                                                                                                %
\includegraphics[width=8.0cm]{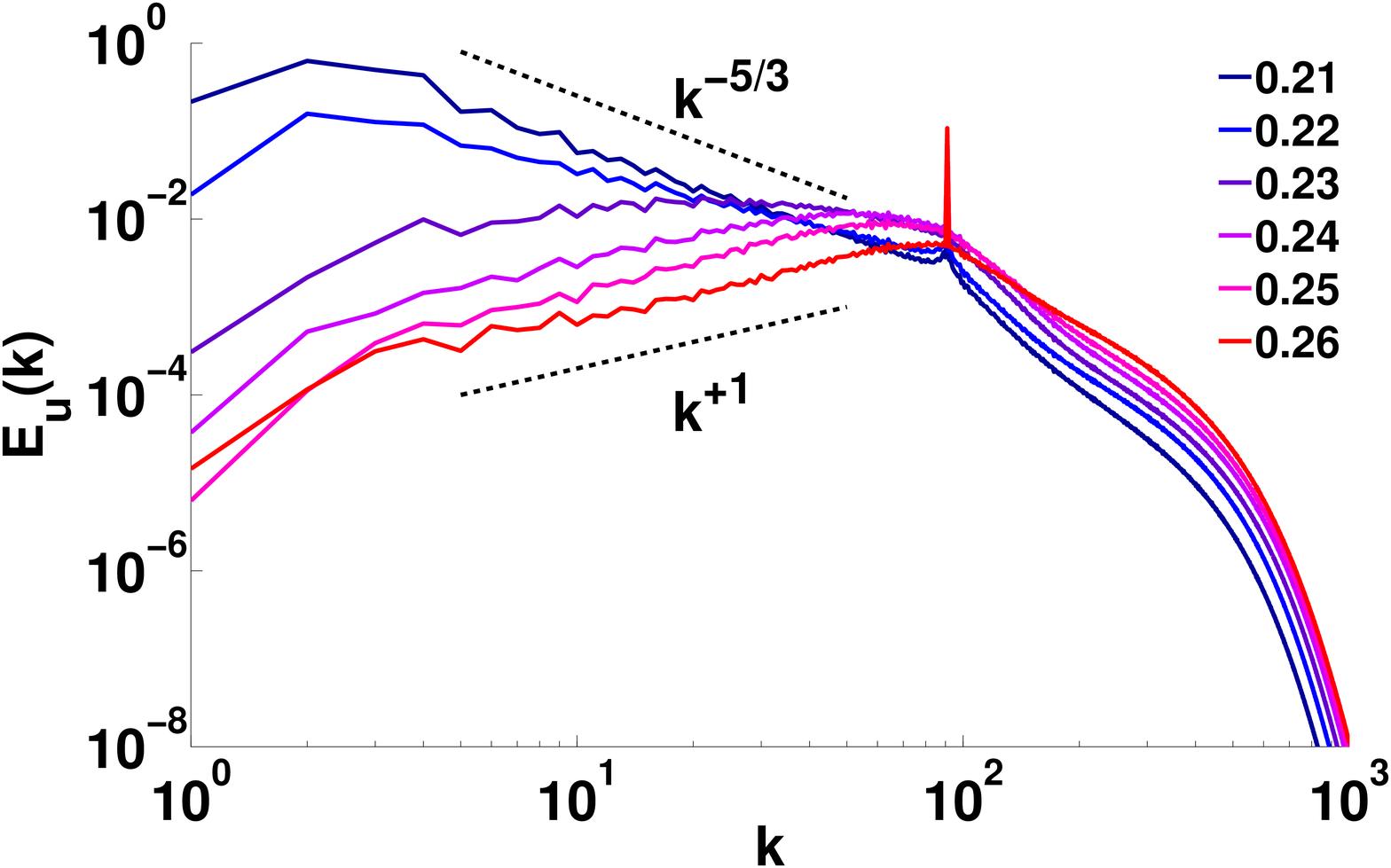}                                                                                    %
\includegraphics[width=8.0cm]{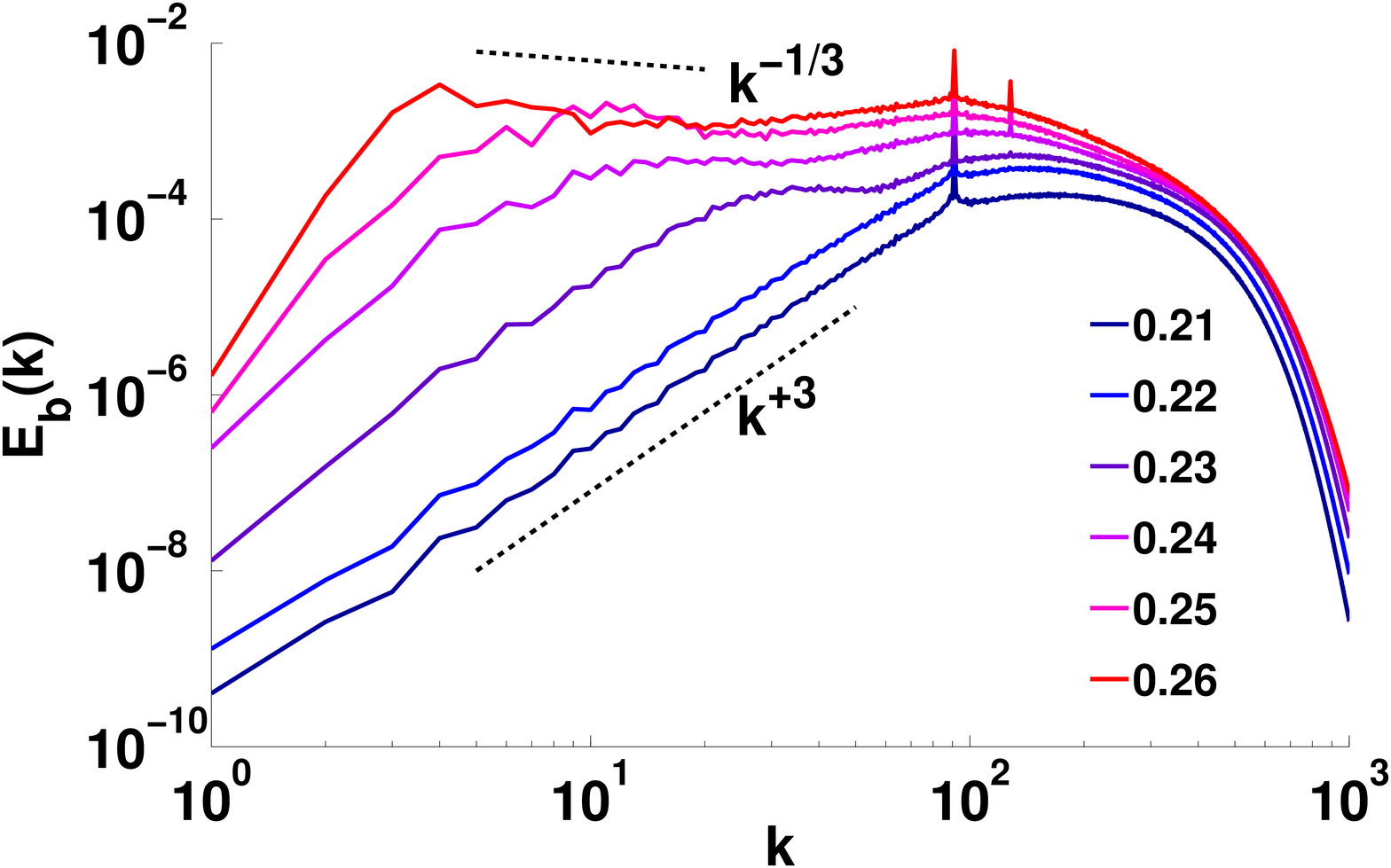}                                                                                    %
\caption{\label{fig6} Kinetic energy spectra (top panel) and magnetic energy spectra (bottom panel)                           %
          for  $k_fL=64$ and different values of $\mu$ varying from 0.21 to 0.26.}                                            %
\end{figure}

Finally, we note that the effect of $\mu$ on the distribution of energy in scale space.
Dimensional phenomenological arguments predict that the energy spectra $E_u$ and $E_b$ of the two fields
will follow different power laws in the two extreme limits. In the large scales ($k<k_f$) for weak magnetic forcing 
we expect the scaling $E_u(k)\sim k^{-5/3}$ for the kinetic energy spectrum and $E_b\sim k^{3}$ if we assume equipartition
of $A$ among all Fourier modes. For strong magnetic forcing we expect the scaling $E_u\sim E_b\sim k^{-1/3}$ due to the 
constant flux of $A$. These exponents however have been criticized before in the literature \cite{Pandit} and have been shown
to be sensitive to dissipation parameters. Figure \ref{fig6} shows the kinetic energy spectra and the magnetic energy spectra 
for different values of the parameter $\mu$ varying from $\mu=0.21$ to $\mu=0.26$ and $k_fL=64$.
It is clear that as $\mu$ crosses the critical value $\mu_c$ the slope of the 
kinetic energy spectrum varies from a value close to $-5/3$ to a value that could be interpreted as $+1$ implying equipartition 
of kinetic energy in all modes. The slope of the magnetic energy spectrum on the other hand decreases from the positive $+3$ value to 
a value close to $-1/3$.
Here, thus, we give an alternative interpretation of the variable exponent that has been measured in the literature for
2D-MHD that the exponent can vary due to the transition from 2D-HD to 2D-MHD. We note that such variation
of the spectral exponent of the inverse cascade have been observed also in rotating turbulence \cite{Rot}.
Of course we note that we have only limited inertial range and what appears as a variable spectral index could in fact be
a smooth transition at some transition wavenumber from the $k^{-5/3}$ regime to the $k^{-1/3}$ regime. 

Concluding, we have demonstrated that the transition from 2D-HD to 2D-MHD by varying the magnetic forcing amplitude 
has a critical behavior. We expect that this is not a unique property of 2D-MHD but can also be observed in 
some of the other systems that were mentioned in the introduction. As far as we know there is no quantitative theory that
describes these transitions. Thus they pose new questions and open new venues for the studies of out of 
equilibrium systems and turbulence, to which the 2D-MHD system can be an ideal ``{\it fruit-fly}" model.

{\it Aknowledgements. }
This work was granted access to the HPC resources of GENCI-CINES (Project No.x2014056421)
and MesoPSL financed by the Region Ile de France and the project Equip\@Meso (reference ANR-10-EQPX-29-01). 
KS acknowledges support from the LabEx ENS-ICFP: ANR-10-LABX-0010/ANR-10-IDEX-0001-02 PSL.
SJB acknowledges support from the College of Natural Sciences Summer Research Abroad Scholarship, University of Texas at Austin.




\end{document}